# Transparent Recovery of Dynamic States on Constrained Nodes through Deep Packet Inspection

Girum Ketema Teklemariam, Floris Van den Abeele, Ingrid Moerman, Jeroen Hoebeke
IDLab - Ghent University - imec
iGent Tower - Department of Information Technology
Technologiepark-Zwijnaarde 15, B-9052 Ghent, Belgium
{firstname.lastname}@ugent.be

*Abstract*— **Many IoT applications make extensive use of constrained devices which are characterized by unexpected failures. In addition, nodes could be put offline temporarily for maintenance (e.g. battery replacement). These incidents lead to loss of dynamic data that is generated due to the interactions between nodes. For instance, configuration settings adjusted by sending PUT request to the sensor will be lost when the node is rebooted. The lost data, which we call dynamic states, leads to erroneous results or malfunctions of the IoT application. In this paper, we introduce an intelligent dynamic state recovery mechanism through deep packet inspection. A State Directory, placed at the gateway, intercepts every communication between an external device and constrained devices and stores (or updates) information that is important to restore dynamic states. When a node reports a reboot, the state directory replays the packets that generated the dynamic state so that all dynamic states are restored. We implemented the solution on a non-constrained device that acts as a gateway to the constrained network and tested the results using Cooja simulator.**

## I. Introduction

IoT applications heavily depend on the interconnection of smart objects through wireless links. The smart objects are characterized by limited capabilities such as memory, power supply, processing power and communication bandwidth. Due to these constraints, the communication network is characterized by unstable links. In combination with the low power aspect, these networks are often referred to as Low-Power and Lossy Network (LLN). Usually, LLNs are connected to external networks, ultimately to the Internet, through gateways. This architecture allows IoT application components to reside entirely in the LLN or to be distributed across different locations (in the LLN, at the gateway and in the cloud). In such models, components of the IoT application residing outside the LLN need to interact with components residing inside. For instance, in a home automation application, a user may adjust the house temperature using his/her smartphone from the Internet. The client is residing outside the LLN while the temperature sensor and actuator are inside.

Different approaches can be used to enable interactions among the different components of the IoT application. One such approach is the use of CoAP-based communication between clients and servers. The Constrained Application Protocol (CoAP) is a light-weight HTTP-like protocol that uses the same GET, PUT, POST and DELETE methods to access and update resource representations hosted on servers [1]. Unlike HTTP, CoAP uses UDP at the transport layer to avoid the overhead introduced by TCP. However, the protocol offers reliability through Confirmable messages. An important extension of CoAP, named Observe [5.2], is also used in IoT applications involving monitoring of resource states. This protocol lets clients register their desire for updated information from the server. After successful registration, the server sends every resource state change to the client until the relationship expires or is proactively terminated by the client. As such, the data collection part of IoT applications can be developed by programming the different components entirely using CoAP and its observe extension.

Such interactions may generate dynamic states that have to be stored in the volatile memory of the constrained device. In addition, other dynamic states such as the list of clients which established observation relationships with the server or bindings [9] are also stored in the volatile memory. All dynamically created states will be lost in case the node reboots for any reason. A reboot can be triggered by a crash or can be the consequence of a firmware update. Such reboots may occur and necessitate the recovery of the application state in order to ensure continuity of the applications outside of the network. This can be achieved by letting the application detect the failure and reinstall the state. This process can be time-consuming and must be managed properly by the applications. Therefore, a fast and automatic recovery mechanism that can deal with crashes in a way that is transparent for the applications is very relevant.

In this paper, a novel approach for crash recovery is introduced. We introduce the concept of a State Directory at the LLN gateway which automatically intercepts and inspects all interactions between clients and servers residing at either side of the gateway. Besides monitoring all application-layer interactions with the LLN, it is able to store all dynamic states generated on the constrained nodes. Upon reception of a start-up message from a node, the state directory replays the requests that generated the dynamic states to recover them on the server. Four types of interactions from external clients that may create dynamic states are intercepted. These are, PUT requests that modify configuration parameters, observation requests, binding requests and dynamic loading requests.

The remainder of the paper is organized as follows. The next section discusses related work in the area of the recovery of nodes and resources. Dynamic states and state recovery is discussed in more detail in Section three. Dynamic state recovery options, including our novel transparent dynamic state recovery mechanism, are explained in section four, while Section five describes the implementation of the proposed solution. Functional and performance evaluation of the proposed mechanism is given in Section six. Section seven concludes the paper.

## II. RELATED WORK

Many IoT applications that involve embedded sensors and actuators rely on LLNs formed by the interconnection of constrained devices. A failure of a constrained device that results in a reboot may affect the entire operation of the application. Several attempts have been made to reduce the impact of such failing nodes on the accuracy of the applications [3]–[5]. Most of these works focus on finding alternative nodes or links in order to avoid malfunctioning. In contrast, our work focuses on regenerating dynamic states on nodes in order to resume their function when they are back online. This way, the IoT application that depends on the installed state is able to continue its operation without taking any specific actions. The aforementioned works focus completely on failed/failing nodes or routes.

Another approach, which has similarities to ours, focuses on storing dynamic states at a location from where it can be restored. The first notable work in this category is given in [6]. The paper presents a sensor network model in which each node stores sensor data locally and provides a database query interface to the data. Each sensor device runs its own database system using Antelope. Antelope provides a dynamic database system that enables runtime creation and deletion of databases and indices. Antelope uses energy-efficient indexing techniques that significantly improve the performance of queries. This technique may be used for storing state information in neighboring constrained devices and can be used to restore the states when the node comes back from failure. However, this approach is different from ours in many aspects. Firstly, the approach needs a database system and an interface to interact with the database on the constrained node, which is expensive for the node. Secondly, the info is stored on a constrained node which is also prone to failure. We select the gateway to store the state information to avoid this issue. Finally, the state information collection is done exclusively using database query interfaces which uses either push or pull methods, whereas our approach is fully transparent to both the client and the server. Yet another notable work is LUSTER [7]. LUSTER presents a hierarchical architecture that includes distributed reliable storage, delay-tolerant networking and deployment time validation techniques. Fault-tolerant storage is provided by discretely listening to sensor node communications without the need of dedicated queries. This is realized through an overlaid, non-intrusive reliable storage layer that provides distributed non-volatile storage of sensor data for online query, or for later manual collection. The storage layer is used for storing sensor data. The permanent storage that is used to store dynamic data is the only similarity of this approach to ours. However, LUSTER uses a storage layer while we use the gateway for permanent storage.

## III. DYNAMIC STATES AND STATE RECOVERY

### A. Dynamic States

IoT applications that make use of embedded web services often involve one or more constrained devices that generate or consume data inside a Low-Power and Lossy Network (LLN). Sensor nodes usually generate data to be consumed by another constrained node or by a non-constrained device residing outside the constrained network. Similarly, actuator nodes consume data that is generated inside the LLN by a sensor node or by a non-constrained device on the Internet. As such, interactions between different components need to take place to realize the desired functionality of IoT applications. In most, if not all, cases, these interactions originate from external devices such as smartphones or monitoring stations. In rare cases, interactions may also originate from the LLN nodes if they are pre-configured to interact with each other. Such interactions may result in the generation of new state information that is either used immediately or stored at the constrained device for future use. For instance, a simple GET request sent to a sensor node from a smartphone usually results in the generation of sensed data and the immediate transmission of responses containing that data. Such information does not need be stored in the LLN as subsequent requests will result in the generation of new data values that will be communicated back. On the contrary, PUT requests that modify the operation threshold of a temperature sensor result in information that needs to be stored locally. We call such information *Dynamic States*. At this juncture, it is important to distinguish between dynamic states and sensor resource states. Sensor resource states are sensor readings exposed as a CoAP resource to be accessed by clients while dynamic states are any set of information generated as a result of interaction between sensors, actuators and other devices and are stored in memory.

Various approaches can be followed to recover lost state information, but all solutions involve storing duplicate information, preferably, on a more permanent or reliable storage. We call these entities *State Directories (SD)*. A State Directory (SD) can be defined to collect and store all dynamic states at a central location. The state directory is then referred and updated regularly by intercepting every potential interaction between devices that may generate or modify dynamic states. The interception may be done by a device that is powerful enough to store and process dynamic states and that has access to the traffic flow without much overhead. Considering this, the LLN gateway may be a good candidate to host the SD and intercept the traffic.

Interactions that result in data that needs to be stored are of great importance for IoT applications. Some of these interactions are described below.

*1) Parameter Modifications at Run-time*

PUT requests sent to nodes in the LLN usually affect some parameters of the node. Such information needs to be stored in memory so that the node can operate as per the new requirement. Examples of such information include alteration of operation thresholds, control parameter modifications and actuations. Every request may change a default value generating a new dynamic state or update the existing one.

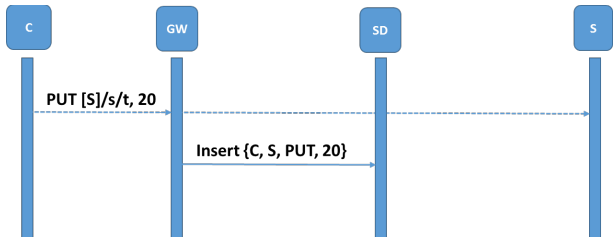

**Figure 1: CoAP Interaction - PUT Request**

Figure 1 shows a PUT request sent to the /s/t resource of the sensor S to change the value to 20. On its way to the destination, the gateway (GW) intercepts the packet and stores the client and server information along with the type of entry and the value in the SD.

*2) Observation Request*

In monitoring applications, clients send observation request to sensors to be registered as observers so that an up-to-date representation is sent to them as soon as it becomes available. After receiving the request, the sensor node stores the details of the client for future notifications. If conditional observe is used, as described in [8], notification criteria will also be stored at the sensor. In addition to this, subsequent notifications may also update the dynamically generated states (Figure 2).

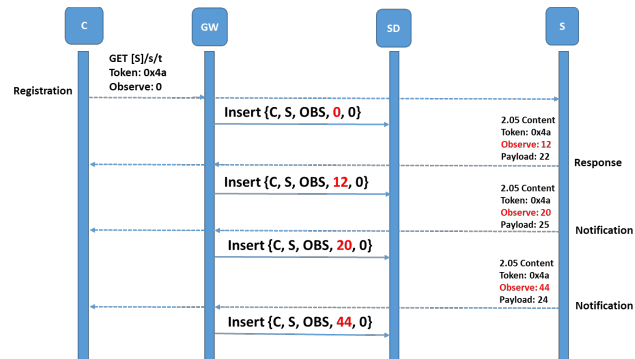

**Figure 2: CoAP Interaction - Observation Request**

Both Observation registration requests and notifications generate dynamic states. In Figure 2, the registration packet result in the generation of a new entry in the SD. In this case, the observe counter, set to 0, and the retransmission counter (also set to 0) along with the client and server information are stored. Subsequent notifications are also shown, updating the observe counter from 0 to 12, 20 and 44, respectively. The retransmission counter is updated only if there are retransmissions that will lead to the cancelation of the relationship when the maximum number of retransmissions is reached.

*3) Flexible Binding Relationship Creation*

A flexible binding relationship is an observation relationship between devices that is established by a third party device [9]. Bindings are established by devices residing outside the LLN and notifications may be sent to nodes in the LLN. Unlike observation relationships, bindings generate new PUT requests instead of responses for the original GET request. The information on how to generate the PUT request needs to be stored by the node.

Only the binding request that is sent from the external device is intercepted by the GW (Figure 3). Binding requests are identified by the BIND_INFO included in the GET request. As the binding information is required to recover the state later on, this information is stored in the state directory. Notifications generated by the sensor and sent to the actuator are not intercepted and have no impact on the information stored in the SD.

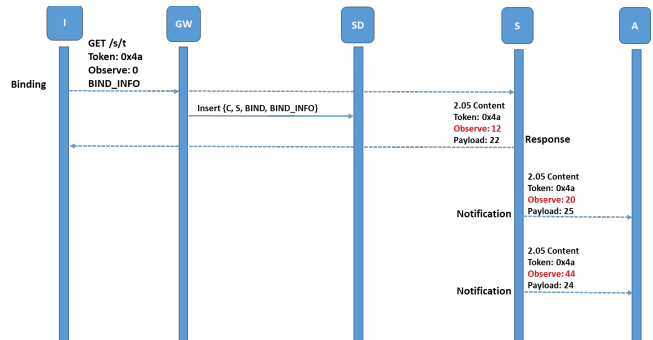

**Figure 3: CoAP Interaction - Binding Request**

*4) Run-time Deployment of Application Code*

Some node architecture allow the deployment of code at runtime [10]. The dynamic code may range from small bug fixes to replacements of a portion of the firmware [11]. Runtime deployment can also be used to provide processing capacity to constrained nodes. Dynamically deployed RESTlets [10] are examples of code fragments that are deployed at runtime. In many cases, the dynamic component is first sent to a permanent storage (e.g. external flash) on the node before it is relocated and loaded into the main memory. Even if the file containing the raw code is in external storage, the relocated, ready-to-use component is stored in memory and can be referred to as dynamic state information. Once loaded into the memory, there are no further interactions that modify the stored information unless it is a new update, which is treated as a new request that replaces the old information.

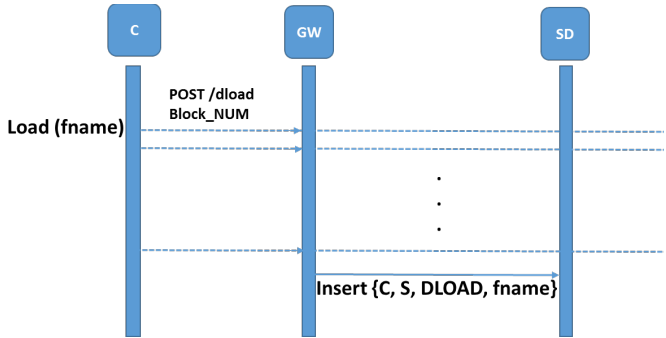

Figure 4: CoAP Interaction – Runtime Deployment

As shown in Figure 4, dynamic modules are transferred using CoAP Block-wise transfer. Once all the blocks have been received by the node, the filename will be stored on the GW together with other relevant information. This will enable the device to dynamically load the module from external memory of the node. Alternatively, we may store the entire dynamic code locally in the SD and replay the transfer of the entire module from the SD.

### B. Recovery of Dynamic States

The states generated through the aforementioned interactions are stored in the memory of the constrained nodes. In this work, we consider state information that is stored in volatile memory and that is lost after rebooting the node. So, in case the constrained nodes go through a power cycle due to an internal error (e.g. software error) or are temporarily put offline for maintenance (e.g. battery replacement), the nodes have lost all dynamic states when they come back online. This situation affects all IoT applications that rely on the previously installed state in that node. A structured and efficient state restoration mechanism is crucial to ensure that all dynamic state information is captured and updated and the latest state is restored when the node comes back to life. Figure 5 shows an example of a structured state restoration process.

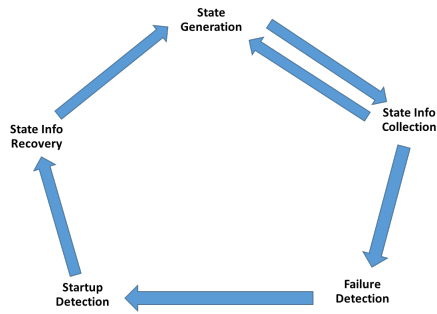

Figure 5: Dynamic State Restoration Cycle

#### 1) State Generation

As explained above, several nodes involved in IoT applications generate dynamic states as a result of interactions with other constrained or non-constrained nodes. Some interactions generate new state information every time, while others continuously update the dynamically created state. For instance, every observation relationship request creates a new dynamic state, whereas notifications generated due to resource state changes update part of the already existing dynamic state (e.g. the latest observe counter).

#### 2) State Information Collection

In order to successfully recover the dynamic states, it is imperative to carefully collect the states as soon as they are generated and/or modified. Some interactions create new states and others modify existing ones. Some interactions may also result in the removal of the dynamic state (For instance, a GET request with the observe option set to 1). Therefore, it is vital to distinguish between these interactions and take the appropriate action. Collection of state information must be done transparently without human intervention and knowledge of the involved parties.

#### 3) Failure Detection

A node might not be available for some time due to physical failure or link failure. In some cases, failure detection is important to defer re-registration (or cancelation) of relationships by clients due to the unavailability of sensors for a short period of time. The best example of such a scenario is an observation relationship established between a client and a sensor node with a fixed Max-Age value. If the node takes more time than the Max-Age value to get back online, the client sends a new GET request to re-register its interest or even a RST to actively cancel the relationship. The re-registration (and possible cancelation) may be avoided in case the failure is detected earlier and an intermediary can respond to the requests in place of the sensor. Early detection may also allow intermediary devices, such as the LLN gateway, to play a role in establishing a separate relationship with replacement devices if the original sensor is gone forever [12]. Once the replacement is put in place having the same IP address, the intermediary may store all states of the old node into the new node so that the transaction continues as before. If such intervention is not required, failure detection is not mandatory, startup detection is.

#### 4) Startup Detection

A node that is not available for some time does not necessarily imply that it has physically failed. Possible link failures on the path between the sensor and the client can break the communication. Timely startup detection is a vital step in state information recovery. A mechanism must be in place to differentiate between physical node failure and link failure to properly recover state information, as link failures typically do not require recovery.

#### 5) State Information Restoration

This is the last step that puts all the dynamic state information back into the memory of the node. Since all transactions are expected to continue as before, we must make sure that all states are restored before the client is aware of the brief absence of the constrained node. In addition, we must make sure that the recovery procedures do not lead to the generation of too much communication creating congestion at specific nodes in the network leading to further timeouts in other communications.

IV. DYNAMIC STATE RECOVERY

As described in the previous section, IoT applications that make use of constrained objects usually depend on dynamically generated states. The loss of such states negatively impacts the performance and/or accuracy of the application. This very idea makes state recovery an important component of IoT applications.

As mentioned earlier, one of the important steps in dynamic state recovery is the collection of the state information in a way that is easy to recover. Dynamic state collection can be done by the node itself using a push mechanism, where the node sends every new or updated state to a pre-configured or negotiated state directory as soon as it is created. The major drawback of this mechanism is the additional overhead it creates in the constrained network, in order to store the states at the SD. Every transaction that gives rise to new or updated dynamic state information results in the generation of a packet towards the SD. This may lead to various problems including an increased power consumption as well as an increased number of packets inside the LLN that results in network congestion. The opposite of this approach is a pull mechanism initiated by the SD. This can be done either by continuous polling or using publish/subscribe mechanisms such as CoAP observe. In both cases, the approach suffers from the same drawback as the push mechanism.

An innovative and less expensive approach is to do the state collection in a way that is transparent to both the client and the server. This can be done by intercepting and inspecting the packets as they traverse the network towards their destination and deciding whether they will result in a new or updated dynamic state. As the interactions that affect the stored dynamic states of a node are known, it is easy to anticipate the change and store the appropriate information in the SD.

Such methods generally work well for unencrypted transactions. Encrypted messages cannot be easily inspected and require another approach. However, this can be overcome by using a a trusted gateway and sensor virtualization as shown in [12]. The next two subsections describe in more detail the transparent state recovery for both unencrypted and encrypted communications, followed by a subsection on alternative approaches.

A. *Transparent Dynamic State Recovery for Unencrypted Communication*

An essential aspect for this approach to work is the location of the state directory. Most interactions, if not all, that alter dynamic states of a constrained device originate from the non-constrained network such as the Internet. Such packets must pass through the LLN gateway before they reach the constrained node. Similarly, responses also go through the gateway on their way to the external device. Moreover, the LLN gateway is a non-constrained device that is always on and can handle the real-time interception and inspection of all packets to and from the LLN. All these features make the LLN gateway an ideal location to place the SD. Our transparent dynamic state recovery solution is based on this basic idea. The gateway intercepts all the traffic between the LLN and the external network and stores all relevant information needed to recreate the dynamic states at a later time when recovery is required (Figure 6). As shown in the figure, a PUT request sent from a smartphone to a constrained node residing in the LLN passes through the gateway, which also houses the state directory. The state directory contains a list of values required to regenerate all dynamic states on the constrained devices.

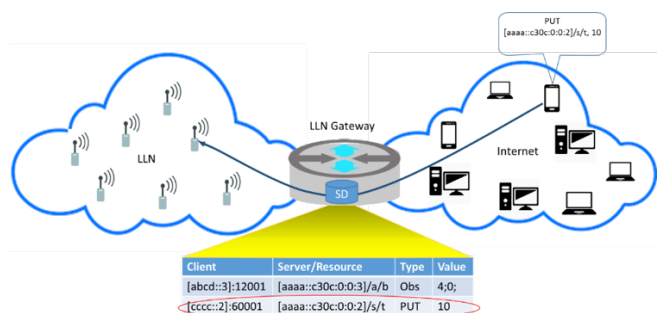

**Figure 6: Placement of State Directory at the LLN Gateway**

Other mechanisms could be incorporated to our solution in order to capture interactions originating and terminating in the LLN, but this is outside the scope of this work. The different steps involved in the transparent dynamic state recovery, as introduced in Figure 5, and how they can be realized are discussed below.

*1) SD – Node Association*

The first step in the transparent dynamic state recovery mechanism is establishing a relationship between the state directory and the sensor nodes inside the LLN. This step is important for subsequent steps too. This can be done in several ways.

**Proactive Registration:** one possible method is to allow explicit registration of nodes upon startup. Whenever a node boots up, it is expected to send a packet to the gateway, which registers the node as a potential node that may store dynamic states. Hereafter, any request-response interaction that may modify the dynamic states on this node will be intercepted. In addition, the registration request can also be used to detect the startup of the node and may trigger the state recovery process. In this case, the packets intercepted will be the ones sent to and from the registered nodes. The main drawback of this approach is that if the registration request is lost in between, the gateway will not be able to know the existence of this node. However, this can be overcome by sending confirmable registration requests.

**Reactive Registration**: by default, the gateway inspects all packets entering the LLN. The moment the first packet is sent towards a node inside the LLN, the node association is created. Further, the packet is inspected and, if needed, the gateway stores the dynamic state provisionally and starts a timer. If an error message is sent back from the LLN stating absence of the node or unwillingness to comply to the request before the timer expires, the provisional information will be removed from the SD. However, if a positive response is intercepted or the timer expires, the provisional status will be changed to permanent. All further packets to and from that node will now be intercepted. This is a completely transparent mechanism where neither the client nor the sensor are aware of the existence of the SD. In addition, no prior registration is required for the operation. However, if an error message is generated and gets lost on its way to the gateway, the gateway will still store a state that does not exist, resulting in inconsistent information. This is the major downside of this method. In addition, we need another mechanism to detect rebooting of the nodes as the approach does not provide an inherent way to detect startups.

**Inference from the LLN Routing Table:** in RPL based networks, the root of the DODAG stores all available nodes in the LLN. The LLN gateway, being closest to the RPL root, may just store all available nodes by collecting the route information from the root node. This can easily be achieved if the root node exposes the route information as a CoAP resource to the gateway and the gateway registers as an observer. The gateway stores all the nodes in the LLN irrespective of their capacity and role and packets to/from these nodes are intercepted. Also nodes that are not involved in any dynamic state generating interaction (such as simple routers) are stored at the SD. This approach suffers from multiple drawbacks. First, re-registration of a node at the DODAG root does not necessarily mean the node is rebooting. When a link to the parent fails, a node chooses a new route and is re-registered at the DODAG root. This information reaches the SD giving it an incorrect information that the node is coming back up from failure. Second, when a node or a link fails, all nodes that use the failed link or node will find an alternate route and notify the DODAG root. Again, this gives inaccurate information to the SD informing it to start the recovery process. Finally, every node sends keep-alives to the root periodically which are treated as updates.

*2) State Information Collection*

State information collection is done to ensure that the correct dynamic state is captured in the state directory. As mentioned before, all requests with the potential of updating a dynamic state must be intercepted so that the SD information is updated. Figure 7 shows how the dynamic state collection works.

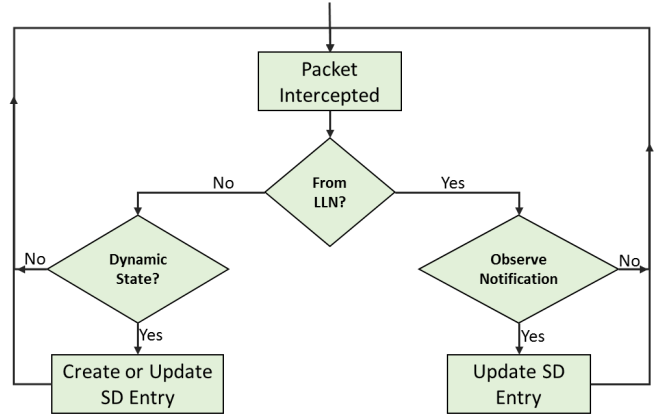

**Figure 7: Dynamic State Information Collection**

Every packet that passes through the gateway is intercepted and different actions are taken based on the origin of the packet. If the packet originates from the LLN, the SD entry associated with the content of the intercepted packet will be updated only if the packet is an observe notification. However, if a request with the potential of creating or modifying a dynamic state on a node in the LLN is intercepted from the Internet, an existing entry will be updated or a new one is created on the SD. As mentioned before, the intercepted requests from the Internet are PUT requests, observe requests, binding Requests and dynamic deployment requests.

A PUT request from the external network triggers the gateway to check the SD for a matching request to the destination. If it exists, the value will be updated. Otherwise, a new entry will be created for the destination. For instance, if a user updates the operating temperature of a thermostat multiple times, the first request creates the information in the SD and subsequent requests will update it.

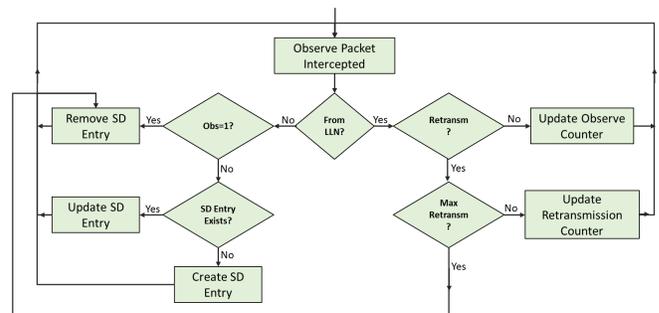

**Figure 8: Observe Request Information Collection**

Observation requests require more thorough examination of the packet before the gateway decides how to update the SD (Figure 8). Observe request from the external network with an observe value different from 1 will establish a new observe relationship between the sender and the LLN node. Accordingly, the gateway creates a new entry in the SD. In case there is already an existing observe relationship between these two parties, this will be indicated by the existence of an entry in the SD. Then, the information will be updated as the same will happen at the LLN node when it receives another observe request from the same client to the same resource. On the other hand, an observe request from the external network with the observe value set to 1, removes an existing relationship and hence is implied by removing an existing entry from the SD. All packets that originate from the LLN that include the observe option are notifications generated as a result of resource state changes affecting an existing observe relationship. So, the gateway captures the packets and updates the current entry in the SD. This is particularly important for two reasons. One, if the node reboots, after successful dynamic state recovery, the observe value used in subsequent notifications must be a continuation of the last notification. Otherwise, the client will ignore the value as obsolete eventually leading to Max-age expiration and, consequently, to re-registration or cancellation of the relationship. The second reason is to capture timeouts. Every now and then, the sensor obliges the client to acknowledge reception of notifications. However, if the client is not available to do so, the sensor retransmits the packets for a fixed number of times and removes the observe relationship in case it fails to get any acknowledgement. This means, the entry should also be removed from the SD. By examining the packet, we detect retransmissions and hence can remove the SD entry after the maximum number of retransmissions is reached.

Binding relationships are established by sending packets from the external network to the LLN. The packets contain the observe option along with the four binding options explained in [10]. Any binding request must be intercepted and the appropriate information stored in the SD. Unlike the regular observe operation, subsequent transactions, except the response to the first request, do not reach the gateway. Therefore, no update will be made for binding relationships after the relationship is established. In order to capture events happening inside the LLN, e.g. cancellation of the binding relationship by either party, different mechanisms may be applied. For instance, the gateway may regularly check the binding directory of the sensor node to see if all bindings are intact.

Requests sent to dynamically loaded RESTlets, can be identified by the block transfer option sent to the dynamic loader resource of the constrained node. As far as storing information in the SD is concerned, we have two options here. The first option is capturing all blocks and storing the dynamic module at the gateway so that recovery can be done by resending the dynamic module to the constrained node. The other option will be storing only the filename and the node's address at the SD. But for this to work, the dynamic module must be available at a permanent storage of the node. Upon reboot, the gateway will send a packet to the node to read and load the dynamic module from the storage.

### 3) Failure Detection

Failure detection is optional and enables the gateway to perform some proxying tasks until the failed node is back online. This can be done by looking at the routing table entries or by periodically polling resources. We do not do failure detection in our implementation.

### 4) Startup Detection

All stored state directory entries must be restored on the corresponding nodes as early as possible in order to avoid the client to miss expected notifications and/or avoid erratic operation of the IoT application. Startup detection triggers restoration of the dynamic states in the node. If nodes proactively register their presence at boot time, this message can be used by the gateway as a mechanism to detect startup. In cases where the node does not get registered at the gateway proactively, other startup detection mechanisms should be in place. One possible method is observing node entries in the LLN routing table at the RPL root coupled with follow-up requests sent to the node. Any change in the LLN is reflected at the RPL root. When a node is not reachable for some time, this information may or may not reach the RPL root immediately. When it comes back, the node tries to become part of the DODAG again and hence the root tries to install this route in the routing table. This information will be communicated to the gateway to indicate that a node has attempted to rejoin the network. However, this does not necessarily mean that the node is recovering from a failure. The node could have been unreachable for various reasons (e.g. mobility). The gateway may send a request to the client to check if an already stored state is available. If the requested dynamic state does not exist or if the value is different, the node is rebooting.

### 5) State Restoration

The final step of the state recovery procedure is state restoration. Once the gateway realizes a node has just finished booting up, it starts the state restoration in a transparent way by replaying the original requests that resulted in the existing dynamic states. For instance, if a PUT request has set the threshold value of a resource, the same request will be sent to the node to restore this value. Similarly, an observe relationship between a device and the node will be restored by sending the same observation request by spoofing the IP address of the originating device. The same works for binding relationships. Recovering dynamic RESTlets may work in either of two ways. If the node has a permanent memory where the dynamic module is stored, the restoration process only takes the filename from the gateway and just does relocation and loading of the module from its local store. Otherwise, the gateway may replay the whole block-wise transfer request to transfer the dynamic modules from the SD.

## B. Transparent Dynamic State Recovery for Encrypted Communication

Secured communication with constrained networks is difficult to be intercepted. In addition, a gateway cannot simply spoof and replay packets. However, [12], gives an innovative way of providing DTLS-based secured communication between LLNs and the Internet by introducing a trusted gateway. The solution exposes a virtual device for every physical device in the LLN at the gateway. This virtual device will minimally expose the same CoAP resources as the physical device. Security between the physical resource and a client in the Internet, is then provided by dividing the connection in two separate secured communication components. The first component securely connects the virtual resources at the gateway and the external client while the second component connects the virtual resource to the physical resource. This way, external networks only see virtualized devices that are perceived as real devices. Every packet addressed to such a virtual node will be terminated at the gateway and a separate packet is sent to the physical node. Responses are also handled in the same manner. This way, it becomes possible to intercept and inspect all packets, even when using DTLS. Of course, the underlying assumption is that the gateway is a trusted device. This can be compared to e.g. SSL/TLS termination in data centers for purposes of load balancing and deep packet inspection.

With this approach, transparent dynamic state recovery can be done, as the interception and inspection of packets takes place at the gateway after decrypting the packets coming from either side of the gateway. Once the inspection is done, the SD entries will be updated or new entries will be created as per the content of the packet. During recovery, the replayed packets from the SD will be encrypted before they are sent to the constrained node.

This mechanism will effectively address the recovery of dynamic states on constrained nodes without affecting the security of the whole system.

## C. Other Dynamic State Recovery Mechanisms

### 1) SD on External Storage

An alternative way of resuming communication after a node comes back from failure is to store a copy of all important dynamic states in external memory (Figure 9). Every time a node boots, it checks its own state directory for the availability of stored states. If there are stored dynamic states, it restores them back into volatile memory before continuing its normal operation. Once normal operation resumes, every change will be stored both in the volatile memory (RAM) and in the storage. In this approach, startup detection is an integral part of the firmware while failure detection is not applicable. State information collection is merely storing new states and modifications in the storage space as a copy. Recovery is the reverse process of copying information from external flash to memory. One of the advantages of this approach is that it does not require additional devices to collect and recover dynamic states. Because of that, no additional packet transmission is required. It is also the fastest way to recover the states. However, it has also its own limitations. First, this approach only works for nodes with external storage which is optional in many smart objects. As external devices are not aware of the process, it is not possible to keep connections alive temporarily in cases where the power cycling process takes longer. Finally, it only works for devices that have been designed in such a way and cannot be retrofitted to legacy devices.

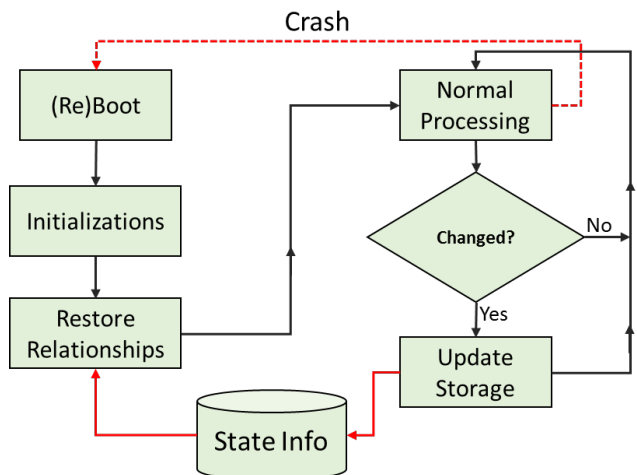

**Figure 9: Dynamic State Recovery with State Directory on the Node**

### 2) SD on Immediate Parent in Multi-Hop Network

Another alternative is to put the state directory at the immediate parent of the node that operates in a multi-hop network. In tree-based multi-hop networks, all communication of a node with the external world goes through a parent node. This makes it easier for the node to store all dynamic states locally to make it available in case of failure of the child node. State information collection may take place transparently by intercepting every packet that passes through the parent node. The parent and child nodes periodically exchange control messages in order to indicate that they are alive. The parent is able to detect the failure of the child node when such messages do not arrive when expected. Upon detecting the failure, the parent node may take actions such as responding for requests on behalf of the child in order to defer re-registration or cancelation of pre-established relationships until the node is back online. The routing control messages that will be sent out upon boot time by a node also reach the immediate parent making its availability known to the parent. The parent may initiate the restoration process as soon as it receives such control messages. Since a third-party is involved in the process, state restoration takes place transparently without the knowledge of both the client and the sensor. In addition, the packet interception at the parent avoids additional packets that might be needed for state collection if the SD was placed elsewhere. Moreover, the parent works on behalf of the sensor until it is back online serving clients with strict deadlines. However, this approach puts a lot of load on the parent node, which is most likely a constrained node itself. In addition, being a constrained device, the parent node may also fail losing all the stored information. Since there is no backup of the parent node, there is no way of putting the states back in the parent node when it resumes functioning after failure. Moreover, topology changes may result in changing parents which actually means losing the stored information.

## V. IMPLEMENTATION

The transparent dynamic state recovery mechanism that has been presented in section 5.1.1 was implemented at the gateway node running CoAP++, an in-house implementation of the CoAP protocol and many additional features using Click Router [13]. The LLN devices are Zolertia Z1 motes [14] running Contiki 2.7 [15] and Erbium [16].

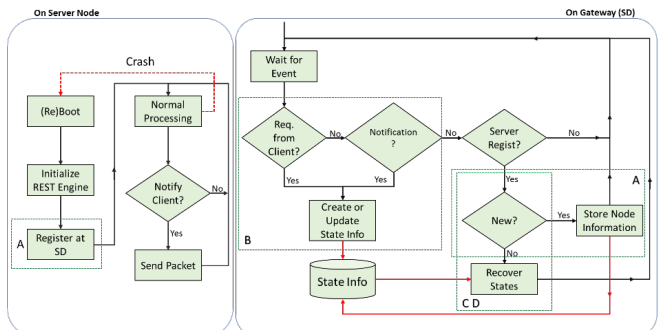

**Figure 10: Implementation of Transparent Dynamic State Recovery**

The SD functionality at the gateway performs the interception just after the IPv6 routing is done. After this step, the destination (sub) network is determined indicating if the transaction is from the LLN to an external network or vice versa. By doing the interception here, we can perform the required processing tasks depending on the origin of the packet. Packets originating from the LLN will be checked for notifications and the exact state information to store (or remove). Figure 10 shows the flow of control to successfully recover lost dynamic states of constrained nodes.

### A. SD – Node Association

We make use of the proactive registration method by nodes to create the association between a node and the gateway. Since the node has earlier knowledge of the gateway, it will just send a registration request to the gateway after initializing its REST engine. Upon receipt of the request, the gateway checks if it has already stored information for that node. The presence of such information ensures that the node is rebooting and triggers recovery. Otherwise, the node's information will be stored in the SD. We use confirmable blocking requests to block all other activities from commencing before the acknowledgement is received from the gateway. This way we can be sure that the association is created properly.

### B. Dynamic State Collection

Dynamic state collection takes place by intercepting all traffic that comes from both sides – the LLN and external network. If the packet is originating from the external network, the state directory will be updated either by creating a new entry or by modifying existing one or even removing entries. Packets originating from the LLN always either modify the existing data (if they are notifications to observers) or remove the entry (if they are retransmissions and max-retransmission is reached).

### C. Startup-Detection

As mentioned above, a proactive registration request sent from an LLN node indicates that the node is booting and also alerts the gateway if there needs to be recovery attempts.

### D. Dynamic State Recovery/Restoration

The decision to restore dynamic states is made by the gateway after it gets a registration request from an LLN node and verifying that the node already has SD entries associated with it. If the stored information is an observe relationship or a PUT request, the gateway replays the original requests by using the original sender's IPv6 address as source address and using the values stored in the SD as required. For instance, the gateway uses the latest observe counter value and the observers IPv6 address when re-establishing the observation relationship and suppresses the response. Restoration of binding relationship requires specifying the binding information along with the observe option value set to 0. The gateway uses one of its own IPv6 addresses as the source address. Finally, to recover dynamic RESTlets, we just specify the name of the file in URI query while sending the recovery information. Upon reception of this packet the node looks for the file in its external memory and reloads it to memory.

## VI. EVALUATION

### A. Functional Evaluation

#### 1) Node-SD Association

**Figure 11: Registration of a node at the gateway**

As explained before, at startup every node sends a CoAP request to a specific resource on the gateway to inform its availability. The message is sent as a blocking request so that every CoAP related operation is blocked until confirmation is received from the gateway. This is required in order to allow the Node-SD association to be in place before further interactions. The SD receives the packet and checks for stored dynamic states for that node and initiates recovery if found (Figure 11). Otherwise, the node is stored as a potential host for dynamic states.

2) *Dynamic State Collection*

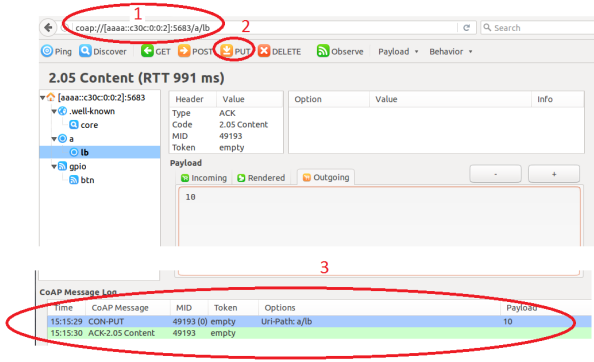

Figure 12: Dynamic State Collection (using PUT method)

Dynamic state collection is done by intercepting all traffic in a way that is transparent to both the client and the server. Figure 12 shows the CoAP Copper Plugin setting the value of the */a/lb* resource to 10 on the node with address *[aaaa::c30c:0:0:2]* through a PUT request (label 1 and 2 on the figure). The server responds that the operation was successful as indicated in label 3. This operation is totally transparent and both the client and the server are unaware of the interception made by the SD. When this request passes through the gateway, the SD intercepts the packet and stores the information locally as shown in Figure 13. The SD entry shows the client address, the server address and the entry type, among other values. Entry Type (ET) 2 means the record is for a PUT request.

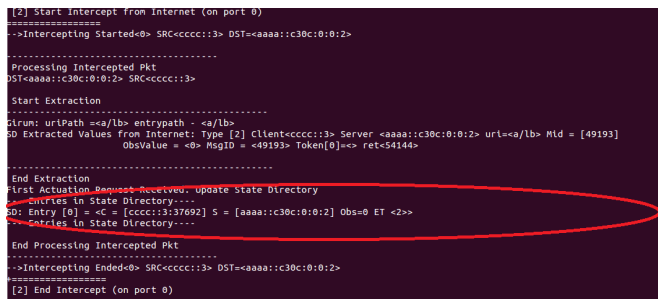

Figure 13: Intercepted PUT Request from Client

3) *Observation Request and Notification Handling*

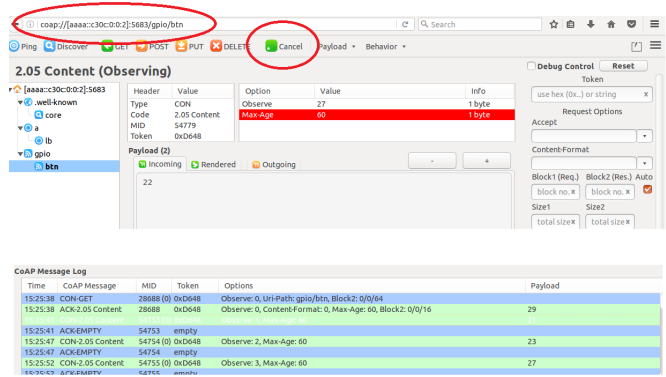

Figure 14: CoAP Observe Request Sent from Copper Client

As explained earlier, observation requests are treated differently as opposed to other intercepted requests. Figure 14 shows Copper sending an observation request to the */gpio/btn* resource on node *[aaaa::c30c:0:0:2]*. Since this is the first observation request from this client, the observe counter is set to 0. This means, the SD needs to store an entry for this request after successful interception (Figure 15).

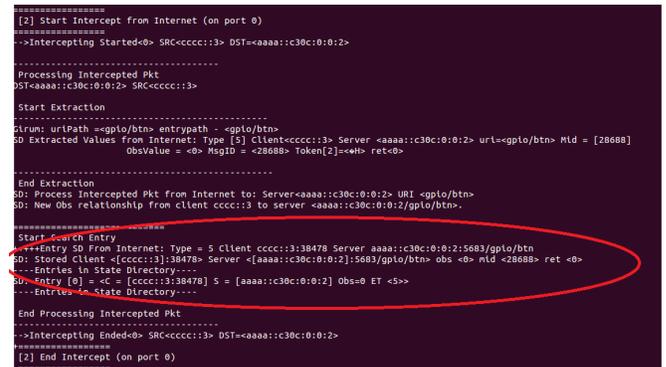

Figure 15: Interception of a New Observe Request at the SD

All notifications sent from the sensor node are intercepted and the SD Entry is updated. When the client wishes to stop the relationship, the RST message will be sent and the entry is removed from the SD.

4) *Dynamic State Restoration*

Figure 16: List of Entries in the SD Before Node [aaaa::c30c:0:0:2]

The restoration process starts as the gateway receives a registration request from a node in the LLN. If entries exist in the SD, the SD initiates the restoration process and generates packets containing the right information to regenerate the dynamic states on the nodes. Figure 16 shows 3 entries in the SD before node [aaaa::c30c:0:0:2] reboots. There are two PUT requests and one Observe request stored in the SD for that node.

Figure 17: SD Containing 3 Records

Upon reception of a registration request from the node, the SD checks the list of entries associated with the registered nodes. Figure 17 shows the 3 records in the SD associated with the rebooted node. The first two records are PUT requests to resources a/lb and a/m, respectively, while the last record is an observation relationship. Accordingly, the SD initiates regeneration of the dynamic states on the node by sending the PUT and Observe requests to the node.

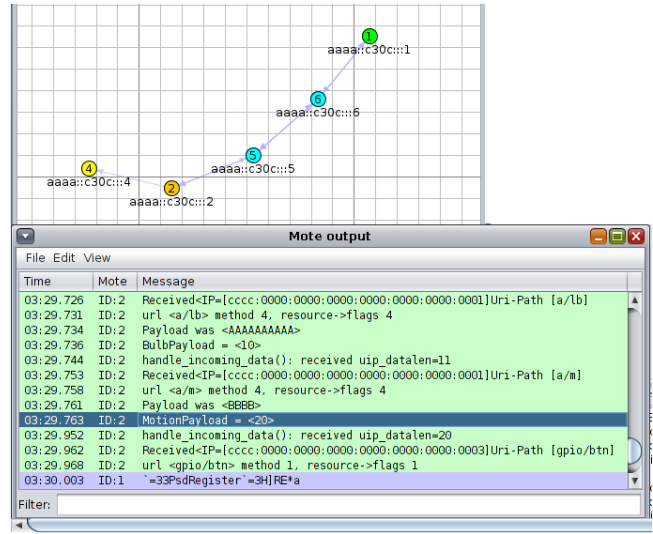

Figure 18: Output of the node after receiving the 3 packets from the SD

Figure 18 shows the outputs of the node on Cooja, illustrating the receipt of the 3 packets while Figure 19 shows the CoAP Message Log in Copper. When we closely look at the output of Figure 18, we see that the sender's address used to regenerate the observation relationship is the address of the client itself, i.e., [cccc::3]. This is important because the server always registers the client as an observer and sends notifications to him. Resumption of normal observation operation can be seen on Figure 19 by simply looking at the message IDs (MID). The change of MID from 12855 to 62713 indicates the re-initialization of the message IDs after reboot.

Figure 19: CoAP Message on Copper Showing Resumption of the Observe Operation

B. *Performance Evaluation*

*1) Delay Introduced by SD-Node Association*

The first step in the transparent crash recovery solution is the association between nodes and the state directory. After rebooting, the node sends a confirmable blocking request to the SD in order to get registered. The SD registers the node and sends a confirmation back. This process introduces delay in the overall crash recovery process. Figure 20 shows the delay introduced due to this process using NullRDC and ContikiMAC as Radio Duty Cycling (RDC) protocols. In both cases, the delay increases with the hop count. This means that larger networks may suffer from longer delays. Yet, the delay for 3 hops is still less than 1 second for ContikiMAC and less than 100ms for NullRDC.

Moreover, the increment is linear and the delay may not be very significant unless the network is too big. However, when we compare the results of ContikiMAC and NullRDC, the difference is quite high even for the same hop count. This is due to the nature of the two RDC protocols. NullRDC keeps the radio on all the time in order to receive packets as soon as the sender attempts to transmit them. This makes all transmission and reception faster but keeping the radio on all the time wastes energy. But ContikiMAC keeps the radio off 99% of the time in order to save energy. Due to this fact, senders have to wait for some time until the sleeping nodes are available making the communication delays higher.

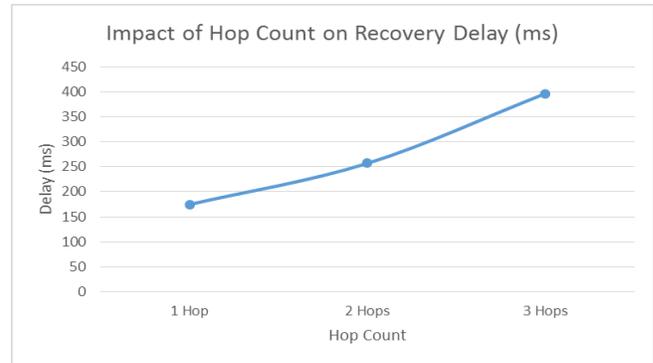

**Figure 21: Impact of Hop Count on Recovery Delay**

*4) Impact of Number of Relationships on a Single Server*

A single server may have multiple dynamic states created as a result of multiple requests from clients. When the node recovers from a crash, every dynamic state has to be recovered. In order to study the impact of multiple states on the recovery time, we sent 3 PUT requests to a single server and measured the recovery delay after the crash. We used ContikiMAC as Radio Duty Cycling (RDC) protocol which lets nodes sleep for most of the time to reduce energy consumed by passive listening. Figure 22 shows that the time required to recover states increases with the number of dynamic states required. The delay is due to the increased number of packets traversing the network, each containing information about a particular dynamic state that will be restored. The impact will be more visible when the number of hops increases because of the number of nodes it traverses to reach the server. When the number of states is a lot more than what we showed here, there is a possibility of congestion while trying to recover the states. In such cases, the SD must have a strategy to inject the recovery requests in the network.

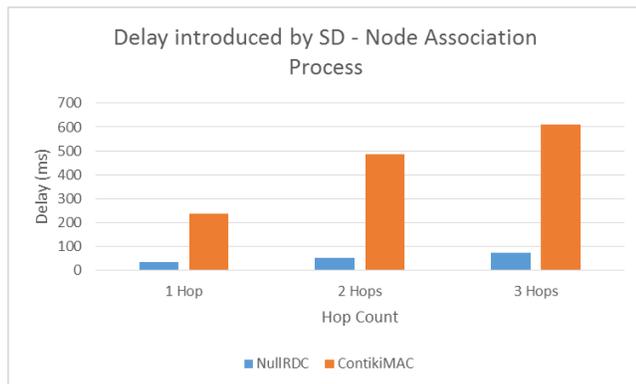

**Figure 20: Delay Introduced by SD-Node Association Process**

*2) Impact of Intercepting Packets*

Every packet that traverses the gateway needs to be intercepted which may introduce delays in the overall recovery process. We measured the arrival time difference by enabling and disabling interception and found out that the impact is minimal (<100ms). This is due to the fact that the interception is being handled by a non-constrained device.

*3) Impact of Routing Hops on the Recovery Process*

Delays are introduced when packets traverse from their source to destination. The delay is especially pronounced in LLNs. Therefore, it is important to study the impact of the distance, expressed in number of hops, between the node and the SD. As expected, the delay increases with the hop count (Figure 21).

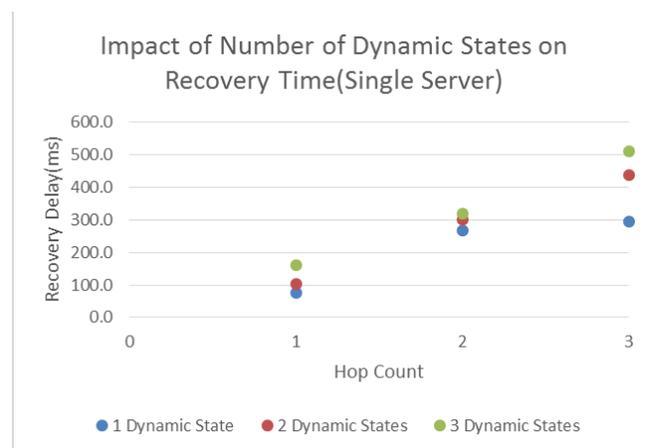

**Figure 22: Impact of Number of Dynamic States that needs to be recovered on recovery time**

## VII. CONCLUSION AND WAY FORWARD

CoAP-based IoT applications may depend on devices located inside a LLN. These devices may reboot for some reason or could be put offline temporarily for maintenance. Dynamic states, values which are created as a result of interaction with other nodes and that are stored in the volatile memory of the nodes, will be lost when the devices are back online. In this paper, we discussed the importance of dynamic state recovery in this context and the possible steps that might be taken to successfully recover the states. In addition, we presented a mechanism for the recovery of these dynamic states in a manner that is transparent to both the client and the server. Our proposed solution intercepts transactions with the potential of creating (and modifying) dynamic states on the nodes at the LLN gateway and stores relevant information in its state directory. The requests that are intercepted are PUT requests that may alter configuration parameters, observation requests, binding requests and dynamic deployment requests. The gateway is selected to host the state directory for two reasons. First, the gateway is a non-constrained device that can handle all interactions. Second, almost all such transactions originate from the outside network and go through the gateway to reach the LLN. The recovery process starts when a node reboots and sends a registration message to the gateway. Upon reception of the message, the gateway checks its state directory for any dynamic state information in the state directory. If a relevant directory entry exists, the gateway sends all the requests that created the dynamic states to the node. The current work focuses on capturing interactions between clients residing outside the LLN and servers inside. In the future, we will work on recovery of dynamic states that are created due to interaction of nodes within the LLN.

## ACKNOWLEDGEMENT

The research leading to these results has received funds from VLIR-UOS in the form of scholarship payment for Girum Ketema Teklemariam.

## REFERENCES

[1] Z. Shelby, K. Hartke, and C. Bormann, "RFC 7252: The Constrained Application Protocol (CoAP)." IETF, pp. 1–112, 2014.

[2] K. Hartke, "RFC 7641: Observing Resources in the Constrained Application Protocol (CoAP)." IETF, pp. 1–30, 2015.

[3] S. Cherrier, Y. M. Ghamri-doudane, S. Lohier, and G. Roussel, "Fault-recovery and Coherence in Internet of Things Choreographies," pp. 532–537, 2014.

[4] A. Akbari, A. Dana, A. Khademzadeh, and N. Beikmahdavi, "Fault Detection and Recovery in Wireless Sensor Network Using Clustering," vol. 3, no. 1, pp. 130–138, 2011.

[5] P. Milano *et al.*, "A Novel Technique for ZigBee Coordinator Failure Recovery and Its Impact on Timing A Novel Technique for ZigBee Coordinator Failure Recovery and Its Impact on Timing Synchronization," no. November, 2016.

[6] N. Tsiftes and A. Dunkels, "A Database in Every Sensor," in *Proceedings of the 9th ACM Conference on Embedded Networked Sensor Systems*, 2011, pp. 316–332.

[7] L. Selavo *et al.*, "LUSTER : Wireless Sensor Network for Environmental Research," in *Proceedings of the 5th international conference on Embedded networked sensor systems*, 2007, pp. 103–116.

[8] G. Teklemariam, J. Hoebeke, I. Moerman, and P. Demeester, "Facilitating the creation of IoT applications through conditional observations in CoAP," *EURASIP J. Wirel. Commun. Netw.*, vol. 1, no. 1, 2013.

[9] G. Teklemariam, F. Van den Abeele, I. Moerman, P. Demeester, and J. Hoebeke, "Bindings and RESTlets: A Novel Set of CoAP-Based Application Enablers to Build IoT Applications," *Sensors (Basel).*, vol. 16, no. 8, 2016.

[10] G. K. Teklemariam, F. Van Den Abeele, P. Ruckebusch, I. Moerman, and P. Demeester, "Dynamic Deployment of RESTlets on Constrained Devices (unpublished)."

[11] P. Ruckebusch, E. De Poorter, C. Fortuna, and I. Moerman, "GITAR : Generic extension for Internet-of-Things ARchitectures enabling dynamic updates of network and application modules," *Ad Hoc Networks*, vol. 0, pp. 1–25, 2015.

[12] F. Van Den Abeele, T. Vandewinckele, J. Hoebeke, I. Moerman, and P. Demeester, "Secure communication in IP-based wireless sensor networks via a trusted gateway Secure communication in IP-based wireless sensor networks via a trusted gateway," no. October, 2015.

[13] E. Kohler, R. Morris, B. Chen, J. Jannotti, and M. F. Kaashoek, "The Click Modular Router," in *Proceedings of the 17th Symposium on Operating Systems Principles*, 1999, pp. 217–231.

[14] Zolertia, "Z1 Datasheet," pp. 1–20, 2010.

[15] T. V. Adam Dunkels, Bj¨orn Gr¨onvall, "Contiki - a Lightweight and Flexible Operating System for Tiny Networked Sensors," in *29th Annual IEEE International Conference on Local Computer Networks*, 2004.

[16] M. Kovatsch, S. Duquennoy, and A. Dunkels, "A low-power CoAP for Contiki," *Proc. - 8th IEEE Int. Conf. Mob. Ad-hoc Sens. Syst. MASS 2011*, pp. 855–860, 2011.